\documentclass[sigconf]{acmart}

\usepackage{balance} 

\usepackage{xspace}
\usepackage{amsmath}  

\newcommand{\fullonly}[1]{}
\newcommand{\shortonly}[1]{#1}

\newcommand{\ieeeonly}[1]{}
\newcommand{\lncsonly}[1]{}
\newcommand{\articleonly}[1]{}
\newcommand{\acmonly}[1]{#1}
\newcommand{\svonly}[1]{}

\newcommand{\myparagraph}[1]{\paragraph*{#1}}

\newcommand{\mycomment}[1]{}

\articleonly{
\usepackage{setspace}
\setstretch{1.1}
\usepackage{fullpage} 
}

\articleonly{}
\lncsonly{}
\acmonly{}

\newcommand{\tuple}[1]{\langle #1\rangle}
\newcommand{\mean}[1]{\left[ \! \left[ #1 \right]\! \right]}

\newcommand{\intersect}{\cap}

\newcommand{\ourlang}{ORAL\xspace}
\newcommand{\cm}{{\it CM}}
\newcommand{\om}{{\it OM}}
\newcommand{\type}{{\rm type}}
\newcommand{\stype}{{\rm sType}}
\newcommand{\scond}{{\rm sCond}}
\newcommand{\rtype}{{\rm rType}}
\newcommand{\rcond}{{\rm rCond}}
\newcommand{\con}{{\rm con}}
\newcommand{\acts}{{\rm acts}}
\newcommand{\Act}{{\it Act}}
\newcommand{\nav}{{\rm nav}}
\newcommand{\val}{{\it val}}

\newcommand{\op}{{\it op}}
\newcommand{\Rules}{{\it Rules}}
\newcommand{\spa}{{\it AU}}

\newcommand{\union}{\cup}

\newcommand{\set}[1]{\{#1\}}

\newcommand{\uncovsp}{{\it uncovAU}}

\newcommand{\Qpol}{Q_{\rm pol}}

\newcounter{lnum}

\newcommand{\mul}{{\rm multiplicity}}

\newcommand{\simplifyrules}{{\rm simplifyRules}}
\newcommand{\wsc}{{\rm WSC}}
\newcommand{\far}{{\rm FAR}}
\newcommand{\frr}{{\rm FRR}}
\newcommand{\ID}{{\rm ID}}

\newcommand{\simplifyrulesL}{\hyperlink{simplifyrules}{\simplifyrules}}

\begin{document}

\acmonly{
\copyrightyear{2019} 
\acmYear{2019} 
\setcopyright{acmcopyright}
\acmConference[SACMAT '19]{The 24th ACM Symposium on Access Control Models and Technologies}{June 3--6, 2019}{Toronto, ON, Canada}
\acmBooktitle{The 24th ACM Symposium on Access Control Models and Technologies (SACMAT '19), June 3--6, 2019, Toronto, ON, Canada}
\acmPrice{15.00}
\acmDOI{10.1145/3322431.3325106}
\acmISBN{978-1-4503-6753-0/19/06}
}

\newcommand{\thanksText}{This material is based on work supported in part by %
    NSF Grants %
    CNS-1421893, 
    and CCF-1414078, 
    ONR Grant N00014-15-1-2208, 
    and DARPA Contract FA8650-15-C-7561. 
}

\title{
Efficient and Extensible Policy Mining for Relationship-Based Access Control
\ieeeonly{\thanks{\thanksText}}\lncsonly{\thanks{\thanksText}}\articleonly{\thanks{\thanksText}}
}

\ieeeonly{\author{
\IEEEauthorblockN{Thang Bui, Scott D. Stoller, and Hieu Le}
\IEEEauthorblockA{Department of Computer Science, Stony Brook University, USA}}}

\articleonly{\author{Thang Bui and Scott~D.~Stoller and Hieu Le\\
    Department of Computer Science, Stony Brook University, USA}}

\lncsonly{\author{Thang Bui \and Scott~D.~Stoller \and Hieu Le}
  \institute{Department of Computer Science, Stony Brook University, USA}}

\acmonly{
\author{Thang Bui}
\affiliation{
\institution{Stony Brook University}}
\email{thang.bui@stonybrook.edu}
\author{Scott~D.~Stoller}
\affiliation{
\institution{Stony Brook University}}
\email{stoller@cs.stonybrook.edu}
\author{Hieu Le}
\affiliation{
\institution{Stony Brook University}}
\email{hle@cs.stonybrook.edu}
}


\svonly{\author{Thang Bui \and Scott~D.~Stoller \and Hieu Le}

\institute{T. Bui \and S. D. Stoller \and H. Le \at Stony Brook University, USA\\ \email{stoller@cs.stonybrook.edu}}

\date{Received: date / Accepted: date}}

\newcommand{\abstracttext}{ %
Relationship-based access control (ReBAC) is a flexible and expressive framework that allows policies to be expressed in terms of chains of relationship between entities as well as attributes of entities.  ReBAC policy mining algorithms have a potential to significantly reduce the cost of migration from legacy access control systems to ReBAC, by partially automating the development of a ReBAC policy.  Existing ReBAC policy mining algorithms support a policy language with a limited set of operators; this limits their applicability.

This paper presents a ReBAC policy mining algorithm designed to be both (1) easily extensible (to support additional policy language features) and (2) scalable. The algorithm is based on Bui et al.'s evolutionary algorithm for ReBAC policy mining algorithm.  First, we simplify their algorithm, in order to make it easier to extend and provide a methodology that extends it to handle new policy language features.  However, extending the policy language increases the search space of candidate policies explored by the evolutionary algorithm, thus causes longer running time and/or worse results.  To address the problem, we enhance the algorithm with a feature selection phase. The enhancement utilizes a neural network to identify useful features.  We use the result of feature selection to reduce the evolutionary algorithm's search space.  The new algorithm is easy to extend and, as shown by our experiments, is more efficient and produces better policies. 
}

\acmonly{
\begin{abstract}
\abstracttext
\end{abstract}}



\acmonly{}

\acmonly{\thanks{\thanksText}}

\maketitle
\ieeeonly{
\begin{abstract}
\abstracttext
\end{abstract}}
\lncsonly{
\begin{abstract}
\abstracttext
\end{abstract}}
\articleonly{
\begin{abstract}
\abstracttext
\end{abstract}}
\svonly{
\begin{abstract}
\abstracttext
\end{abstract}}


\section{Introduction}
\label{sec:intro}

In {\it relationship-based access control} (ReBAC), access control policies are expressed in terms of chains of relationships between entities.  This increases expressiveness and often allows more natural policies.  High-level access control policy models such as attribute-based access control (ABAC) and ReBAC are becoming increasingly widely adopted, as security policies become more dynamic and more complex.  ABAC is already supported by many enterprise software products, using a standardized ABAC language such as XACML or a vendor-specific ABAC language.  Forms of ReBAC are supported in popular online social network systems and are starting to emerge in other software systems as well.

High-level policy models such as ReBAC allow for concise and flexible policies and promise long-term cost savings through reduced management effort.  However, the initial cost of developing a ReBAC policy to replace an existing lower-level policy can be a significant barrier to adoption of ReBAC.  {\em Policy mining} algorithms promise to drastically reduce this cost, by automatically produce a ``first draft'' of a high-level policy from existing lower-level data.  There is a substantial amount of research on role mining, surveyed in \cite{mitra2016survey,das2018}, and a small but growing literature on ABAC policy mining \cite{xu15miningABACShort,xu14miningABAClogs,medvet2015,mocanu2015,sparselogs2018,iyer2018}, surveyed in \cite{das2018}.

Bui et al. proposed a problem of ReBAC policy mining \cite{bui17mining,bui19mining}: given information about subjects, resources, and other objects, and the set of currently granted permissions, find a ReBAC policy that grants the same permissions using high-level rules. For realistic datasets, the search space of possible policies is enormous.  In traditional ABAC languages, such as XACML, each expression involves at most one attribute dereference, e.g., subject.department, where subject (the entity making the access request) is a User, and User.department is the user's department affiliation.  In ReBAC, an expression may contain a path involving several attributes, and the search space grows exponentially in the path length.  For example, Bui et al.'s ReBAC policy for healthcare contains the expression resource.record.patient.treatingTeam, where resource (the entity to which access is requested) is an Entry in an electronic health record, Entry.record is the HealthRecord containing the entry, HealthRecord.patient is the patient that the health record is for, and Patient.treatingTeam is the clinical team treating the patient.

Bui et al. also proposed a ReBAC policy language, called \ourlang (Object-oriented Relationship-based Access-control Language), which formulates ReBAC as an object-oriented extension of ABAC.  Rules are built from {\em atomic conditions}, which involve a single object (e.g., subject.department='ComputerScience') and {\em atomic constraints}, which relate two objects (e.g., subject.department=resource.department).  Relationships are expressed using attributes that refer to other objects, and {\em path expressions} are used in conditions and constraints to follow chains of relationships between objects.

Bui et al. developed two ReBAC policy mining algorithms \cite{bui17mining,bui19mining}.  Their {\em greedy algorithm} uses heuristics to construct and then generalize candidate rules, attempts to merge and simplify the candidate rules, and then selects the best ones to include in the final policy.  Their {\em Evolutionary Algorithm} (EA) re-uses part of the greedy algorithm to construct some candidate rules, and then uses evolutionary searches starting from the candidate rules to find improved rules to include in the final policy.

This paper proposes a new algorithm for ReBAC policy mining that builds on and improves Bui et al.'s EA.  We build on EA because it performs better overall than their greedy algorithm in their experiments, and because it is easier to extend, as discussed below.  Our algorithm, called FS-SEA*, has two phases, enclosed in an outer loop that iterates until a complete policy has been generated (the ``*'' in the name indicates this iteration).  The first phase, called {\em feature selection} (FS) identifies a relatively small set of atomic conditions and atomic constraints that are likely to be ``useful'', i.e., to appear in the desired ReBAC rules; we call these {\em useful features}.  Our feature selection algorithm is based on machine learning, specifically, neural networks (NNs).  We chose NNs over other AI classification methods due to their flexibility and their scalability to high-dimensional data and large datasets.  NNs are good at implicitly learning high-level features, including the interactions between multiple input features, whereas other classifiers such as SVM often require manual feature engineering to achieve high classification accuracy.  The second phase, called {\em simplified evolutionary algorithm} (SEA), is a simplified version of EA that is also modified to consider only rules built from the useful features identified in the first phase.  Our feature selection phase could easily be added to other policy mining algorithms as well.


FS-SEA*'s advantages over EA include extensibility, efficiency, and effectiveness, as discussed next.

\myparagraph{Extensibility}  Extending FS-SEA* to handle extensions to the policy language is much easier than extending EA, especially when extending the language with additional operators that can appear in constraints.  EA's first phase (and Bui et al.'s greedy algorithm) contains {\em rule generalization} step that attempts to generalize initially constructed candidate rules by removing conditions and adding constraints.  This step computes a set of candidate atomic constraints to possibly include in the rule, and then performs a relatively costly (worst-case exponential) search to find the optimal subset of them to include in the rule.  Increasing the set of possible constraints, by adding more operators, significantly increases the cost of this step.  We simplify EA by removing the rule generalization step.  Instead, SEA relies on the evolutionary search to find appropriate generalizations of the candidate rules in the initial population.  Note that this step cannot simply be removed from the greedy algorithm, which contains no other mechanism that can serve the same purpose.

To show that SEA works as well as EA---in other words, that this simplification has no ill effects---we ran SEA and EA on several policies used in \cite{bui19mining}, and found that the algorithms give similar results (see Section \ref{sec:evol-alg}).

To show that FS-SEA* is easy to extend, we provide general guidelines for extending it, and illustrate them by supporting two constraint operators not considered by Bui et al.: set-equality (i.e., equality between sets; the equality operator in \cite{bui19mining} is applied only to primitive values, not sets, in constraints) and subseteq.

\myparagraph{Efficiency and Effectiveness}

Evolutionary algorithms are based on randomized search and therefore intrinsically involve a trade-off between efficiency (running time) and effectiveness (quality of results).  It is usually possible to get better results, at the expense of longer running time, simply by increasing the limit on the number of search steps.  

To show that feature selection significantly improves the efficiency-effectiveness trade-off (or ``cost-benefit ratio''), we developed a pseudorandom synthetic policy generator to produce a variety of policies that use the two new operators mentioned above as well as the existing operators in \ourlang, ran SEA and FS-SEA* on the synthetic policies, with the same limit on the number of search steps in both algorithms, and found that FS-SEA* was {\bf significantly faster} and achieved {\bf significantly better results}.

Both of these benefits resulted from feature selection successfully focusing the evolutionary search on the most promising part of the search space, preventing it from wasting time exploring less promising parts.  It is obvious how this leads to better results.  It also leads to smaller running time, because SEA generates lower-quality rules that, on average, each cover fewer of the permissions granted by the given low-level access control policy, hence it needs to generate more rules, and this takes longer (a more detailed explanation is in Section \ref{sec:evaluation-results}).  As the policy language is further extended, and the search space grows further, it is expected that the benefits of using feature selection to focus the search will also increase. 

To show the benefits of performing feature selection multiple times, we also ran experiments comparing FS-SEA* with a simpler version called FS-SEA1 that omits the outer loop mentioned above, and calls FS and SEA only once.  We found that the use of iteration in FS-SEA* yields slight to moderate improvements in the results, at the expense of a small increase in running time.

\section{Policy Language}
\label{sec:language}

We adopt Bui et al.'s \ourlang (Object-oriented Relationship-based Access-control Language) \cite{bui19mining}, with some extensions, as our policy language.  We give a brief overview of \ourlang and refer the reader to \cite{bui19mining} for details.  We also describe two new constraint operators, namely subseteq and set equality, that we add to \ourlang as illustrative language  extensions.

Our main contribution---namely, our feature selection technique and accompanying empirical demonstration of its benefits---treats policy language constructs as features in a generic way, and can easily handle additional extensions to \ourlang and be used with other ABAC or ReBAC policy languages.  We include an overview of \ourlang here to make this paper more self-contained and, more importantly, to emphasize \ourlang's expressiveness.  The expressiveness of the policy language, and the consequent vastness of the search space of possible policies, makes the policy mining problem especially challenging and drives the need for new techniques to improve the scalability and effectiveness of policy mining algorithms.

A {\em ReBAC policy} is a tuple $\pi=\tuple{\cm, \om, \Act, \Rules}$, where $\cm$ is a class model, $\om$ is an object model, $\Act$ is a set of actions, and $\Rules$ is a set of rules.

A {\em class model} is a set of class declarations. 
Each field has a {\em type}, which is a class name or ``Boolean'', and a {\em multiplicity}, which specifies how many values may be stored in the field and is ``one'' (also denoted ``1''), ``optional'' (also denoted ``?''), or ``many'' (also denoted ``*'', meaning any number).  Boolean fields always have multiplicity 1.  Every class implicitly contains a field ``id'' with type String and multiplicity 1.  A {\em reference type} is any class name (used as a type).  Bui et al. allow inheritance between classes.  We do not consider inheritance in this paper but plan to consider it in future work.

An {\em object model} is a set of objects whose types are consistent with the class model and with unique values in the id fields.  
Let $\type(o)$ denote the type of object $o$. 
The value of a field with multiplicity ``many'' is a set.  The value of a field with multiplicity  ``optional'' may be a single value or the placeholder $\bot$ indicating absence of a value.


A {\em path} is a sequence of field names, written with ``.'' as a separator.  
A {\em condition} is a set, interpreted as a conjunction, of atomic conditions.  
An {\em atomic condition} is a tuple $\tuple{p, \op, \val}$, where $p$ is a non-empty path, $\op$ is an operator, either ``in'' or ``contains'', and $\val$ is a constant value, either an atomic value or a set of atomic values.  
For example, an object $o$ satisfies $\tuple{{\rm dept.id}, {\rm in}, \{{\rm CompSci}\}}$ if the value obtained starting from $o$ and following (dereferencing) the dept field and then the id field equals CompSci.
For readability, we usually write conditions with a logic-based syntax, using ``$\in$'' for ``in'' and ``$\ni$'' for ``contains''.  For example, we may write $\tuple{{\rm dept.id}, {\rm in}, \{{\rm CompSci}\}}$ as ${\rm dept.id} \in \{{\rm CompSci}\}$.  We may use ``='' as syntactic sugar for ``in'' when the constant is a singleton set; thus, the previous example may be written as dept.id=CompSci.

A {\em constraint} is a set, interpreted as a conjunction, of atomic constraints.  Informally, an atomic constraint expresses a relationship between the requesting subject and the requested resource, by relating the values of paths starting from each of them.  An {\em atomic constraint} is a tuple $\tuple{p_1, \op, p_2}$, where $p_1$ and $p_2$ are paths (possibly the empty sequence), and $\op$ is one of the following four operators: equal, in, contains, supseteq, subseteq.  
Implicitly, the first path is relative to the requesting subject, and the second path is relative to the requested resource.  The empty path represents the subject or resource itself.  For example, a subject $s$ and resource $r$ satisfy $\tuple{{\rm specialties}, {\rm contains}, {\rm topic}}$ if the set $s$.specialties contains the value $r$.topic.

 For readability, we usually write constraints with a logic-based syntax,  using ``$=$'' for ``equal'' and ``$\supseteq$'' for ``supseteq'',
 and we prefix the subject path $p_1$ and resource path $p_2$ with ``subject'' and ``resource'', respectively.  For example, $\tuple{{\rm specialties}, {\rm contains}, {\rm topic}}$ may be written as ${\rm subject.specialties} \ni {\rm resource.topic}$.

A {\em rule} is a tuple $\langle$ {\it subjectType}, {\it subjectCondition}, {\it resourceType}, {\it resourceCondition}, {\it constraint}, {\it actions} $\rangle$, where {\it subjectType} and {\it resourceType} are class names, {\it subjectCondition} and {\it resourceCondition} are conditions, {\it constraint} is a constraint, {\it actions} is a set of actions.  
A rule must satisfy several well-formedness requirements \cite{bui19mining}.
For a rule $\rho=\tuple{st, sc, rt, rc, c, A}$, let $\stype(\rho)=st$, $\scond(\rho)=sc$, $\rtype(\rho)=rt$, $\rcond(\rho)=rc$, $\con(\rho)=c$, and $\acts(\rho)=A$.

For readability, we may prefix paths with ``subject'' or ``resource'', to indicate the object from which the path starts. 
For example, the e-document case study \cite{bui19mining,decat14edocShort} involves a large bank whose policy contains the rule: A project member can read all sent documents regarding the project.  This is expressed as $\langle\,$Employee, subject.employer.id = LargeBank, Document, true, subject.workOn.relatedDoc $\ni$ resource, \{read\}$\rangle$, where Employee.workOn is the set of projects the employee is working on, and Project.relatedDoc is the set of sent documents related to the project.



The {\em type of a path} $p$ (relative to a specified class), denoted $\type(p)$, is the type of the last field in the path.\fullonly{  The {\em multiplicity of a path} $p$ (relative to a specified class), denoted \hypertarget{mul}{$\mul(p)$}, is one if all fields on the path have multiplicity one, is many if any field on the path has multiplicity many, and is optional otherwise.}  Given a class model, object model, object $o$, and path $p$, let \hypertarget{nav}{$\nav(o,p)$} be the result of navigating (a.k.a. following or dereferencing) path $p$ starting from object $o$. 
The result might be no value, represented by $\bot$, an atomic value, or\shortonly{ (if a field in $p$ has multiplicity many)}\fullonly{ (if $p$ has multiplicity many)} a set of values.
This is like the semantics of path navigation in UML's Object Constraint Language\fullonly{ (\url{http://www.omg.org/spec/OCL/})}.

An object $o$ {\em satisfies} an atomic condition $c=\tuple{p, \op, \val}$, denoted $o\models c$, if $(\op={\rm in} \land \nav(o,p) \in \val) \lor (\op={\rm contains} \land \nav(o,p) \ni \val)$.\shortonly{  Objects $o_1$ and $o_2$ {\em satisfy} an atomic constraint $c=\tuple{p_1, \op, p_2}$, denoted $\tuple{o_1,o_2} \models c$, is defined in a similar way.}\fullonly{  The {\em meaning} of a condition $c$ relative to a class $C$, denoted $\mean{c}_C$ is the set of instances of $C$ (in the implicitly given object model) that satisfy $c$.  A condition $c$ {\em characterizes} a set $O$ of objects of class $C$ if $O$ is the meaning of $c$ relative to $C$.

Objects $o_1$ and $o_2$ {\em satisfy} an atomic constraint $c=\tuple{p_1, \op, p_2}$, denoted $\tuple{o_1,o_2} \models c$, if $(\op={\rm equal} \land \nav(o_1,p_1) = \nav(o_2,p_2)) \lor (\op={\rm in} \land \nav(o_1,p_1) \in \nav(o_2,p_2)) \lor (\op={\rm contains} \land \nav(o_1,p_1) \ni \nav(o_2,p_2)) \lor (\op={\rm supseteq} \land \nav(o_1,p_1) \supseteq \nav(o_2,p_2))$.

} An {\em SRA-tuple} is a tuple $\tuple{s, r, a}$, where the ``subject'' $s$ and ``resource'' $r$ are objects, and $a$ is an action, representing (depending on the context) authorization for $s$ to perform $a$ on $r$ or a potential request to perform that access.
An SRA-tuple $\tuple{s, r, a}$ {\em satisfies} a rule $\rho=\langle st, sc, rt,$ $rc, c, A\rangle$, denoted $\tuple{s, r, a} \models \rho$, if
$\type(s)=st \land s\models sc \land \type(r)=rt  \land r\models rc  \land \tuple{s,r}\models c \land a \in A$.  The {\em meaning} of a rule $\rho$, denoted $\mean{\rho}$, is the set of SRA-tuples that satisfy it.
The {\em meaning} of a ReBAC policy $\pi$, denoted $\mean{\pi}$, is the union of the meanings of its rules.


\section{Problem Definition}
\label{sec:problem}

We adopt Bui et al.'s definition of the ReBAC mining problem.  We repeat the core parts of the definition here, and refer the reader to \cite{bui19mining} for additional description and details.

An {\em access control list (ACL) policy} is a tuple $\tuple{\cm, \om, \Act, \spa}$, where $\cm$ is a class model, $\om$ is an object model, $\Act$ is a set of actions, and $\spa\subseteq \om\times \om \times \Act$ is a set of SRA tuples representing authorizations.  Conceptually, $\spa$ is the union of the access control lists.

An ReBAC policy $\pi$ is {\em consistent} with an ACL policy $\langle\cm, \om,$ $\Act,$ $\spa\rangle$ if they have the same class model, object model, actions, and $\mean{\pi} = \spa$.

Among the many ReBAC policies consistent with a given ACL policy $\pi_0$, the most desirable ones are those that satisfy the following two criteria.  One criterion is that the ``id'' field should be avoided when possible, because policies that use this field are (to that extent) identity-based, not attribute-based or relationship-based.  Therefore, the ``id'' field should be used only when necessary, i.e., only when every ReBAC policy consistent with $\pi_0$ uses it.  The other, more generic, criterion is that the policy should have the best quality as measured by a given policy quality metric $\Qpol$, expressed as a function from ReBAC policies to the natural numbers, with small numbers indicating high quality.  This is natural for metrics based on policy size, which is the most common choice.

The {\em ReBAC policy mining problem} is: given an ACL policy $\pi_0=\langle \cm, \om,$ $\Act, \spa\rangle$ and a policy quality metric $\Qpol$, find a set $\Rules$ of rules such that the ReBAC policy $\pi=\tuple{\cm, \om, \Act, \Rules}$ is consistent with $\pi_0$, uses the ``id'' field only when necessary, and has the best quality, according to $\Qpol$, among such policies.

The policy quality metric that our algorithm aims to optimize is {\em weighted structural complexity} (WSC), a generalization of policy size first introduced for RBAC policies \cite{molloy10mining} and later extended to ReBAC \cite{bui19mining}.  Minimizing policy size is consistent with 
usability studies showing that more concise access control policies are more manageable \cite{beckerle13formal}.  WSC is a weighted sum of the numbers of primitive elements of various kinds that appear in a rule or policy. WSC is defined bottom-up.   The $\wsc$ of an atomic condition $\tuple{p, \op, \val}$ is $|p| + |\val|$, where $|p|$ is the length of path $p$, and $|\val|$ is 1 if $\val$ is an atomic value and is the cardinality of $\val$ if $\val$ is a set.  The $\wsc$ of an atomic constraint $\tuple{p_1, \op, p_2}$ is $|p_1|+|p_2|$.  The WSC of a rule $\rho$, denoted $\hypertarget{wscRule}{\wsc(\rho)}$, is the sum of the WSCs of the atomic conditions and atomic constraints in it, plus the cardinality of the action set (more generally, it is a weighted sum of those numbers, but we take all of the weights to be 1). The WSC of a ReBAC policy $\pi$, denoted $\hypertarget{wscPol}{\wsc(\pi)}$, is the sum of the $\wsc$ of its rules. 


\section{Feature Selection (FS)}
\label{sec:fs}

A {\em feature} is a subject atomic condition, resource atomic condition, or atomic constraint satisfying the user-specified limits on lengths of paths in conditions and constraints.  We define a mapping from feature vectors to Boolean labels: given an SRA-tuple $\tuple{s,r,a}$, we create a feature vector (i.e., a vector of the Boolean values of features evaluated for subject $s$ and resource $r$) and map it to true if the SRA-tuple is permitted (i.e., is in $\spa$) and to false otherwise.  We represent Booleans as integers: 0 for false, and 1 for true.  We train a NN to learn this classification (labeling) of feature vectors.  We then analyze the weights learned in the NN to quantify how ``useful'' (important) each feature is in determining the NN's output.  We then rank the features according to their usefulness and classify the highest-ranked features as ``useful features''.


We decompose the problem by learning useful features separately for each tuple consisting of a subject type, a resource type, and an action.  Specifically, we learn a separate neural network $NN_{C_s, C_r, a}$ to classify SRA-tuples with subject type $C_s$, resource type $C_r$, and action $a$.  We do this for each $\tuple{C_s, C_r, a}$ such that $\spa$ contains some SRA-tuple with a subject of type $C_s$, a resource of type $C_r$, and action $a$.  The inputs to $NN_{C_s, C_r, a}$ are limited to the features appropriate for subject type $C_s$ and resource type $C_r$, e.g., the path in the subject condition starts with a field in class $C_s$.   The set of labeled feature vectors used to train $NN_{C_s, C_r, a}$ contains an element generated from each possible combination of a subject of type $C_s$ (in the given object model) and resource of type $C_r$.  We modify SEA so that, when initially constructing rules that cover SRA-tuples with a particular subject type $C_s$, resource type $C_r$, and action $a$, it uses only the features classified as useful based on the edge weights in $NN_{C_s, C_r, a}$.


This decomposition is justified by the fact that all SRA-tuples authorized by the same rule must contain subjects with the same subject type and resources with the same resource type.  A rule can authorize SRA-tuples with different actions, since the last component of a rule is a set of actions.  SEA initially learns rules containing a single action; at the end, it attempts to merge similar rules with different actions into a single rule authorizing multiple actions.

As an optimization, we discard a feature if it has the same truth value in all of the labeled feature vectors used to train a NN; for example, if all instances of some type $C$ in the given object model have the same value for a field $f$, then atomic conditions on field $f$ will be discarded.

We also detect sets of \textit{equivalent features}, which are features that have same truth value in all feature vectors labeled true and used to train a NN. For each set of equivalent features, we keep only the features with the lowest WSC, and discard the rest.  The discarded features should not be used in the mined policy, because any rule using them would have unnecessarily high WSC.



\subsection{Neural Network Architecture and Training}
\label{sec:fs:nn}

\begin{figure}[htb]
  \centering
\includegraphics[width=0.45\textwidth]{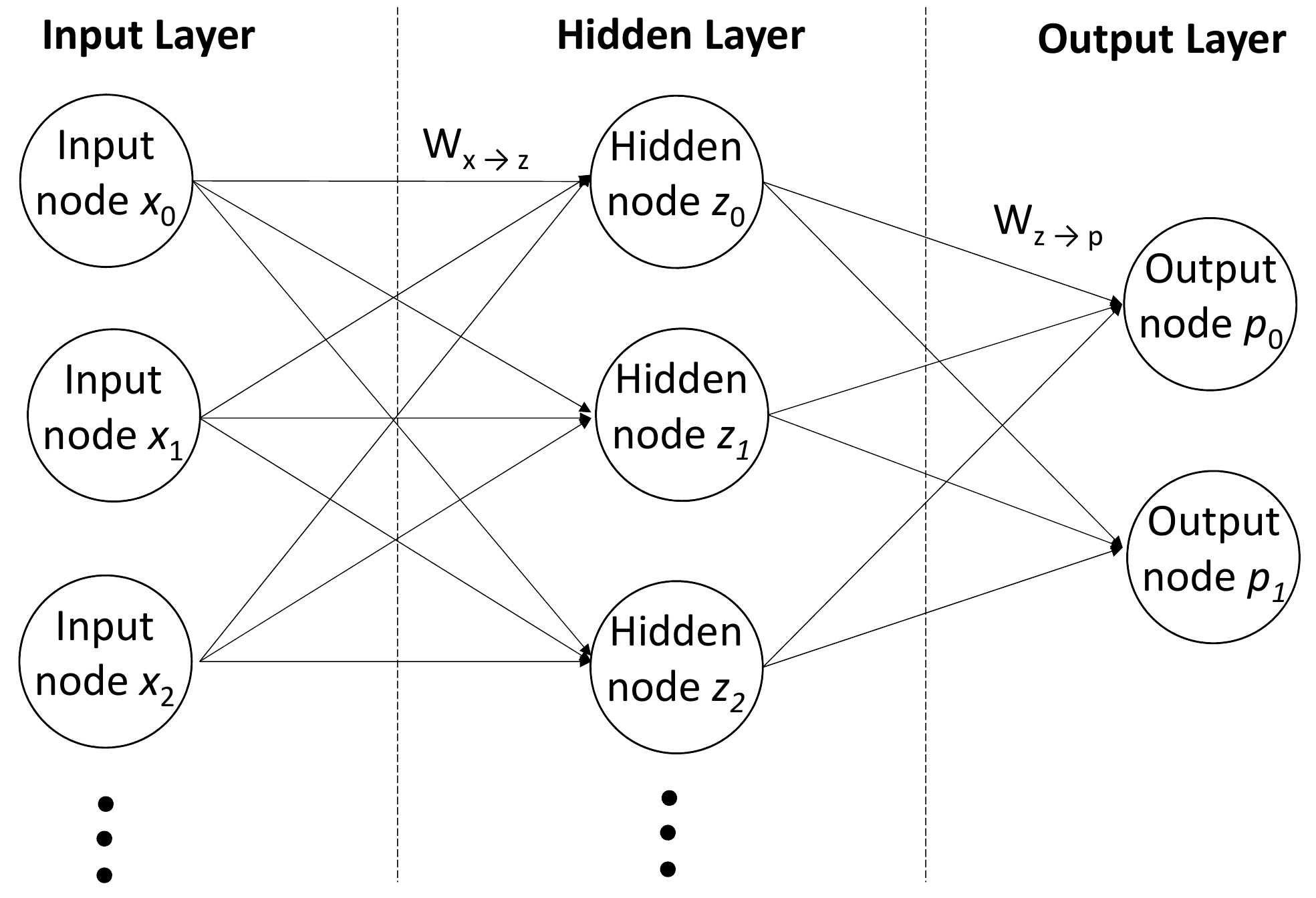}
  \caption{Architecture of the 2-layer neural network.}
  \label{fig:nn}
\end{figure}

We use a 2-layer NN architecture.  The input layer contains an input node $x_i$ corresponding to each feature in the feature vector.  The hidden layer contains hidden nodes $z_i$ that help the NN represent more complex, nonlinear classifiers.  The hidden nodes use the popular rectified linear unit (ReLU) activation function \cite{relu2015}: 
\begin{equation}
     z_i = \max(0, a_i)
\end{equation}
where the input $a_i$ to hidden node $z_i$ is a weighted sum of the values $x_j$ of the input nodes: $a_i = \sum_{j} w_{x_j \rightarrow z_i} x_j$, where $w_{x_j \rightarrow z_i}$ is the weight of the connection from input node $x_j$ to hidden node $z_i$.  
Finally, the output layer contains two output nodes $p_i$, each associated with an access decision: $p_0$ with deny, and $p_1$ with permit. 
The activation function for the output nodes is softmax, a common choice for NNs used as classifiers:
\begin{equation}
  p_i = \frac{e^{b_i}}{e^{b_0} + e^{b_1}}
\end{equation}
where the input $b_i$ to output node $p_i$ is a weighted sum of the values $z_j$ of the hidden nodes: $b_i = \sum_{j} w_{z_j \rightarrow p_i} z_j$, where $w_{z_j \rightarrow p_i}$ is the weight of the connection from hidden node $z_j$ to output node $p_i$.  Note that $b_0$ is considered the (main) input to $p_0$, even though $b_1$ is also used in the computation of $p_0$, because $b_1$ is merely used for normalization.  The outputs $p_0$ and $p_1$ are estimates of the probability that the input should be classified as denied or permitted, respectively.  The classification decision is the one with higher probability: permit iff $p_1 > p_0$, and deny iff $p_0 > p_1$.  If $p_0=p_1$, the feature vector is unclassified.

The training objective is to minimize the cross-entropy for the two classes (i.e., decisions): 
\begin{equation}
\label{eq:crosse-entropy}
   L = \sum_{n}-w_0 \cdot y_0^n \cdot \log(p_0^n) - w_1 \cdot y_1^n \cdot \log(p_1^n)
\end{equation} 
where $n$ indexes the training inputs, each consisting of a feature vector $v^n$ and its label $\ell^n$, and where $\tuple{y_0^n, y_1^n}$ is $\tuple{1, 0}$ if $\ell^n = 0$ and is $\tuple{0, 1}$ if $\ell^n = 1$.  This objective function (a.k.a. loss function) reaches its absolute minimum, which is zero, when $\tuple{y_0^n, y_1^n} = \tuple{p_0^n, p_1^n}$ for all $n$, i.e., when the NN's outputs exactly match the label for each feature vector instance.  We define the weights $w_i$ in the loss function in a way that compensates for the fact that our training data contains many fewer positive instances than negative instances: $w_1 = (N-N_1)/N$ and $w_0 = 1 - w_1$, where $N$ is the number of training inputs to this NN, and $N_1$ is the number of training inputs to this NN with label $\ell^n=1$.

We train the neural network using standard backpropagation, with ADADELTA \cite{adadelta2012}, an adaptive learning rate method, as the optimization algorithm.  For ADADELTA's hyperparameters, we use the default values in the PyTorch implementation $\epsilon=10^{-6}$ and $\rho = 0.9$.  We continue training until all inputs are correctly classified (and no inputs are unclassified) or a specified limit $N_{tr}$ on the number of training iterations is reached.  We originally used stochastic gradient descent as the optimization algorithm, but it required manual tuning of the learning rate and still took longer to learn an accurate classifier.

We experimented other NN architectures and training algorithm.  We tried other activation functions, such as LeakyReLU for the hidden layer, and sigmoid for the hidden layer and output layer. For the optimization algorithm when training the network with backpropagation, we tried Stochastic Gradient Descent (SGD).  With these variations, it took longer to learn an accurate classifier.  We also tried NNs with multiple hidden layers.  These more complex NNs did not help the system learn an accurate classifier faster and would have required a more complex analysis to identify useful features.



\subsection{Useful Feature Selection}
\label{sec:fs:selection}

We rank features using the following score that measures how much each feature contributes to each output, based on the weights on the paths in the NN from the associated input node to each output node.  The score is computed in two steps.

First, compute the contribution of each hidden node $z$ to a permit outcome as $s^1_z = w_{z \rightarrow p_{1}} - w_{z \rightarrow p_{0}}$, and its contribution to a deny outcome as $s^0_z = w_{z \rightarrow p_{0}} - w_{z \rightarrow p_{1}}$.  We use the difference between the weights to the two output nodes, because the difference in the output values determines the outcome (i.e., the classification decision).

Second, for each input node $x$ and each possible outcome, compute the sum, over hidden nodes $z$ with positive weights from $x$, of the product of the weight from $x$ to $z$ and the hidden node's score for that outcome computed in step 1; we use the product (instead of sum) of those weights to reflect the behavior of the product in the definition of $a_i$.  Thus, $s^0_x = \sum_{z} \max(0, w_{x \rightarrow z}) \cdot s^0_z$, and  $s^1_x = \sum_{z}\max(0, w_{x \rightarrow z}) \cdot s^1_z$.  We found that using $\max(0, w_{x \rightarrow z})$ to effectively drop terms with negative $w_{x \rightarrow z}$ gives better results than using the absolute value $|w_{x \rightarrow z}|$ or simply $w_{x \rightarrow z}$.  Intuitively, this is because the input's contribution to increasing the value of one output node relative to the other is most important to the outcome.

 As useful features, we select $N_{uf} = F_u \cdot N_f$ features, where $N_f$ is the total number of features, and $F_u$ (a parameter of the algorithm) is the fraction of features to be selected as useful.  We initially experimented with the straightforward approach of selecting the $N_{uf}$ features with the highest values of $s^1_x$ as useful features.  However, this approach sometimes missed some desired useful features.  For example, when a desired feature had the form $p$=True, where $p$ is a path, the learned NN might give $p$=True high weight to $s^1$, or the NN might give $p$=True lower weight to $s^1$ and compensate by giving its ``complementary'' feature $p$=False higher weight to $s^0$; both NNs can classify feature vectors accurately, and the objective function does not  prefer one of them over the other.  To ensure the desired feature is categorized as useful regardless of which NN is learned in this and similar situations, we select the $\frac{1}{3} N_{uf}$ features with the highest values of $s^0_x$ and the $\frac{2}{3} N_{uf}$ features with the highest values of $s^1_x$.  In addition, whenever a feature of the form $p$=True (or $p$=False) is added to the set of useful features, the complementary feature $p$=False (or $p$=True, respectively) is also added to the set of useful features.


\myparagraph{Other approaches to feature ranking}

We also experimented with other approaches to rank features by importance.  We tried the Grad-CAM technique \cite{grad-cam} when experimenting with NNs with multiple hidden layers.  It is designed for convolutional neural networks but can be applied to any NN, since it only requires computing the gradients of the NN.  We also tried an approach inspired by the backward-propagation based explanation methods in \cite{lemna}, which rank features by their contribution in determining the NN's output for a given feature vector.  We adopted this approach by evaluating the contribution of each feature for each positive-labeled feature vector. In particular, we used the main feature ranking method described above, except that for each positive-labeled feature vector, we ranked only the features whose value is True in that feature vector. A set of useful features is then determined for each positive-labeled feature vector. The final set of useful features is the union of these sets.  In our experiments, the approach we adopted works at least as well as any these alternative approaches.

\myparagraph{Extensibility}

When extending the policy language with new operators, the feature selection 
algorithm requires no changes, because it treats all features in a generic way. It requires only the ability to evaluate a feature to a truth value for a given subject and object.

\subsection{Example}
\label{sec:fs:example}

We illustrate feature selection on a fragment of {\em Electronic Medical Record (EMR)} sample policy, a ReBAC policy based on the EBAC policy in \cite{bogaerts15entity} and available at \cite{fssea-software}.  It controls access by physicians and patients to electronic medical records, based on institutional affiliations, patient-physician consultations (each EMR is associated with a consultation), supervisor relationships among physicians, etc.  To keep this example small, we consider here only one rule in the policy: A physician can create a medical record associated with a consultation if the physician is not a trainee, the consultation is with the physician, and the patient of the consultation is registered at the hospital with which the physician is affiliated.  This is expressed as $\rho=\langle$Physician, subject.isTrainee=false, Consultation, true, subject = resource.physician $\land$ subject.affiliation $\ni$ resource.patient.registrations, \{createMedicalRecord\}$\rangle$.

The relevant run of the feature selection algorithm is for subject type Physician, resource type Consultation, and action createMedicalRecord.  For this pair of types, with the path length limits in \cite{bui19mining}, there are 474 possible features.  We focus on the features $f_1$, $f_2$, and $f_3$ that appear in $\rho$ and (for illustrative purpose) one feature $f_4$ that does not, where
\begin{eqnarray*}
  f_1 &=& \mbox{subject.isTrainee = false}\\  
  f_2 &=& \mbox{subject = resource.physician}\\
  f_3 &=& \mbox{subject.affiliation $\in$ resource.patient.registrations}\\
  f_4 &=& \mbox{subject.consultations.records $\subseteq$ resource.records}
\end{eqnarray*}
The equivalent features optimization discards features such as atomic condition ``subject.consultations.id $\ni$ \{consultation24-1\}'' and atomic constraint ``subject.consultations.records $\supseteq$ resource.records''.
The discarded features are not included in the above count of possible features.

Each feature vector corresponding to a subject $s$ and resource $r$ has the form $\langle$ $s$.id, $r$.id, value of $f_1$, value of $f_2$, value of $f_3$, value of $f_4$, $\ldots$ ; label$\rangle$. For example, the feature vector for the permitted SRA-tuple $\langle$doc12, consultation2-2, createMedicalRecord $\rangle$ is $\langle$doc12, consultation2-2, 1, 1, 1, 0, $\ldots$ ; 1$\rangle$.

To illustrate feature ranking, we describe some calculations of $s^1_x$ using edge weights in NN$_{\rm Physician,Consultation,createMedicalRecord}$; calculations of $s^0_x$ are similar.  We first compute the contribution of each hidden node to a permit outcome; for example, for $z_0$,
\[ s^1_{z_0} = w_{z_{0} \rightarrow y_{1}} - w_{z_{0} \rightarrow y_{0}} = 0.0704 - ({-0.06070}) = 0.1311\]
Let $x_i$ be the input node corresponding to feature $f_i$.  By definition, $s^1_x = \sum_{z}\max(0, w_{x \rightarrow z}) \cdot s^1_z$.  For feature $f_2$, we have 
$s^1_{x_2} = \max(0, w_{x_2 \rightarrow z_0}) \cdot s^1_{z_0} + \cdots 
= \max(0, 0.1907) \cdot 0.1311 + \cdots = 0.0250 + \cdots = 2.1028$.
For feature $f_4$, we have 
$s^1_{x_4} = \max(0, w_{x_4 \rightarrow z_0}) \cdot s^1_{z_0} + \cdots 
=\max(0, -0.0236) \cdot 0.1311 +\cdots = 0 + \cdots = 0.0492$.
Based on the computed values of $s^1_x$, $f_2$ is ranked 2 out of 474, and $f_4$ is ranked 347 out of 474, in terms of their contribution to a permit outcome.  With $F_u=0.05$ (as in our experiments with FS-SEA*), $f_2$ is selected as useful, and $f_4$ is not.


\section{Simplified Evolutionary Algorithm}
\label{sec:evol-alg}

Our Simplified Evolutionary Algorithm (SEA) is based on Bui et al.'s evolutionary algorithm (EA) \cite{bui19mining}.   It uses context-free grammar genetic programming (CFGGP) \cite{gramEvolSurvey2010} to search for high-quality ReBAC rules.

ReBAC rules are represented in the algorithm as derivation trees of a context-free grammar (CFG).  The main part of EA is preceded by {\it grammar generation}, which specializes the generic grammar of \ourlang to a specific input. The language of the generated grammar contains rules satisfying the restrictions: (1) constants are limited to those appearing in the object model, (2) class names and field names are limited to those appearing in the class model, (3) paths in conditions and constraints are type-correct, based on the class model, and satisfy the same length limits as mentioned in Section \ref{sec:fs}, and (4) actions are limited to those appearing in the given authorizations.  The grammar generation algorithm pre-computes all atomic conditions and atomic constraints satisfying these restrictions.  

EA's first phase iterates over the given SRA-tuples (authorizations), and uses each of the selected SRA-tuples as the seed for an evolutionary search that adds one new rule to the candidate policy.  Each evolutionary search starts with an initial population containing candidate rules created from a seed SRA-tuple along with numerous random variants of those rules together with some completely random candidate rules, evolves the population by repeatedly applying genetic operators (mutations and crossover), and then adds the highest quality rule in the population to the candidate policy.  Rule quality is measured using the same fitness function $f$ as \cite{medvet2015} (our definition is slightly simplified but equivalent):
$f(\rho) = \tuple{\far(\rho),\frr(\rho),\ID(\rho),\wsc(\rho)}$, where the {\em false acceptance rate} is $\far(\rho)=|\mean{\rho}\setminus\uncovsp|$, the {\em false rejection rate} is $\frr(\rho)=|\uncovsp\setminus\mean{\rho}|$, 
$\uncovsp$ is the subset of $\spa$ not covered by the current candidate policy, and $\ID(\rho)$ equals 2 if the subject condition and resource condition both contain an atomic condition with path ``id'', equals 1 if exactly one of them does, and equals 0 if neither of them does.  The fitness ordering is lexicographic order on these tuples, where smaller is better.  The first phase ends when the candidate policy covers $\spa$.  The second phase improves the candidate rules by further mutating them, and then attempts to simplify each rule and merge similar rules. 

The set of genetic operators used in the search phase contains: (1) single mutation: first, randomly select whether to mutate the subject condition, resource condition, or constraint, then randomly select a non-terminal $N$ in that part of the derivation tree, and then randomly re-generate the subtree rooted at $N$; (2) double mutation: same as single mutation, except, in the first step, choose two out of the three possibilities, and then perform the remaining steps for both of them; (3) action mutation: in the action set component of the rule, randomly add or remove actions that subject $s$ can perform on $r$ according to $\spa$, subject to the restriction that we never remove the action in the seed tuple for this search; (4) simplify mutation: remove one randomly selected atomic condition (from the subject condition or resource condition) or atomic constraint; (5) crossover: randomly select a non-terminal $N$ in the subtree for the subject condition, resource condition, or constraint in one parent, find the same non-terminal in the other parent (if it does not appear, select a different non-terminal in the first parent), and swap the subtrees rooted at those two occurrences of $N$.

The set of genetic operators used in the improvement phase contains: (1) single mutation; (2) double mutation; (3) type+single mutation: randomly select whether to replace the subject type, resource type, or both with their parent types (if those parents exist), apply a single mutation, check whether the resulting rule is well-formed (because the unchanged condition or constraint might be inconsistent with the changed type), and if not, discard it; (4) type+double mutation: same as type+single mutation, except with a double mutation instead of a single mutation.

We describe some of the genetic operators as if they directly manipulate abstract syntax trees, because this description is more intuitive.  However, all genetic operators actually manipulate derivation trees of the generated grammar.

The improvement phase might seem redundant, because it uses essentially the same mutations as the first phase.  The key difference is that, in phase 1, the benefit of a mutation is evaluated by its effect on rule quality, and in phase 2, it is evaluated in the context of the entire candidate policy by its effect on policy quality.  For example, consider a mutation that transforms a candidate rule $\rho$ into $\rho'$, such that $\rho'$ covers fewer SRA-tuples, has lower WSC, and has lower rule quality.  If this mutation occurs in phase 1, $\rho'$ might survive, but it is likely to be discarded, due to its lower rule quality.  If this mutation occurs in phase 2, and if the tuples covered by $\rho$ and not by $\rho'$ are also covered by other rules in the candidate policy, $\rho'$ will definitely replace $\rho$ in the candidate policy, because this change reduces the policy's WSC and does not change the policy's meaning.

The main difference between EA and SEA is that we simplified the step that constructs candidate rules to include in the initial population, by eliminating one of the sub-steps, namely {\it rule generalization}, which is harder to extend to support new language features.  Rule generalization was responsible primarily for selecting a subset of the candidate constraints to include in the candidate rule.  Instead, SEA simply generates a separate candidate rule for each candidate atomic constraint, and relies on the subsequent evolutionary search to generate a rule with a good subset of candidate constraints.

\myparagraph{Extensibility}

SEA can easily be extended to support additional operators in the policy language.  We give generic instructions describing which parts of the algorithm might need to be modified.  There are several of them, but the required changes are straightforward.  (1) Extend the functions that generate candidate rules to support the new operators.  Since our algorithm is based on CFGGP, this basically means extending the context-free grammar to include the new operators.  (2) Extend the function that evaluates conditions and constraints (to obtain their truth value for a given subject and object) to handle the new operators.  (3) Add rule simplification transformations specific to the new operators, if any.  For example, when extending the algorithm to support subseteq and set-equality, we added one simplification: if a constraint contains two atomic constraints on the same paths, one atomic constraint with supseteq operator and the other with subseteq operator, then replace them with a single atomic constraint on the same paths with set-equality operator.  (4) If the new operators may appear in conditions, extend the rule merge function to handle the new operators, by defining an appropriate upper bound function, which is used to replace atomic conditions on the same path in two rules being merged with an equivalent or looser atomic condition. For example, if we allowed subseteq in conditions (our current implementation allows it only in constraints, but we could easily allow it in conditions), then we would define the upper bound of $p \subseteq c_1$ and $p \subseteq c_2$, where $p$ is a path and $c_1$ and $c_2$ are constant sets, to be $p \subseteq c$ where $c = c_1 \union c_2$.



\myparagraph{Efficiency and Effectiveness}

To ensure that the simplification and the extensions do not adversely affect the algorithm's efficiency or effectiveness, we ran EA and SEA on the university, project management, and health care sample policies and the workforce management case study in \cite{bui19mining}.  For all three sample policies, both algorithms produce policies that are identical to the simplified input policies.  For the workforce management case study, both algorithms generate policies with similar quality, as measured by syntactic similarity (defined in Section \ref{sec:evaluation-methodology}) to the simplified input policy.  The two algorithms have similar running times for all of these policies.

\section{Overall Algorithms}
\label{sec:fs-sea}

\myparagraph{FS-SEA1}
Our first algorithm is called Feature Selection---Simplified Evolutionary Algorithm 1 (FS-SEA1).   Its first phase is the feature selection (FS) algorithm in Section \ref{sec:fs} to compute a set of useful features for each $\langle$ subject type, resource type, action $\rangle$ tuple.  Its second phase is SEA, slightly modified to consider only rules built from the appropriate set $UF$ of useful features. Specifically, we modify the function that generates the initial population so that it uses only features in $UF$, and we modify the grammar specialization algorithm to eliminate parts of the grammar corresponding to atomic conditions and atomic constraints not in $UF$.


We noticed in experiments with FS-SEA1 that sometimes, when multiple rules need to be mined for one $\langle$ subject type, resource type, action $\rangle$ tuple, and one of the desired rules covers only a small subset of the relevant SRA-tuples in $\spa$, that rule was sometimes not mined correctly, because the necessary features were not selected as useful.  Since the rule covers a small number of SRA-tuples, the edge weights associated with the features in it are small (though still large enough for the NN to correctly classify the SRA-tuples), and thus those features are given low rankings by our feature selection algorithm.

Our first approach to overcoming this problem was to modify our feature selection algorithm to use L1 regularization or L2 regularization, in the hope that this would boost the ranking of the desired features.  However, this approach was not effective.


Our second approach to overcoming this problem is to perform feature selection multiple times, so that, each time, it can focus on identifying the features useful for covering the remaining uncovered tuples.  This approach was much more successful and is adopted in our second algorithm, described next.

\myparagraph{FS-SEA*} 
Our FS-SEA* algorithm is similar to FS-SEA1, except that it runs multiple iterations of FS and SEA.  In each iteration, the algorithm runs FS and then SEA, adds to the current mined policy $\pi$ only the single highest-quality candidate rule for each $\langle$ subject type, resource type, action $\rangle$ tuple, checks whether the rules in $\pi$ together cover all of $\spa$, and if not, starts another iteration, using $\spa \setminus \mean{\pi}$ as the set of SRA-tuples to cover (except that, when checking whether a candidate rule is valid, SEA still uses all of $\spa$).  This allows each iteration of FS and SEA to focus on the uncovered SRA-tuples.


\myparagraph{Extensibility}

When extending the policy language with new operators, the overall algorithm requires no additional changes beyond the changes to SEA described in Section \ref{sec:evol-alg}.

\section{Evaluation Methodology}
\label{sec:evaluation-methodology}

\begin{figure}[htb]
  \centering
\includegraphics[width=0.48\textwidth]{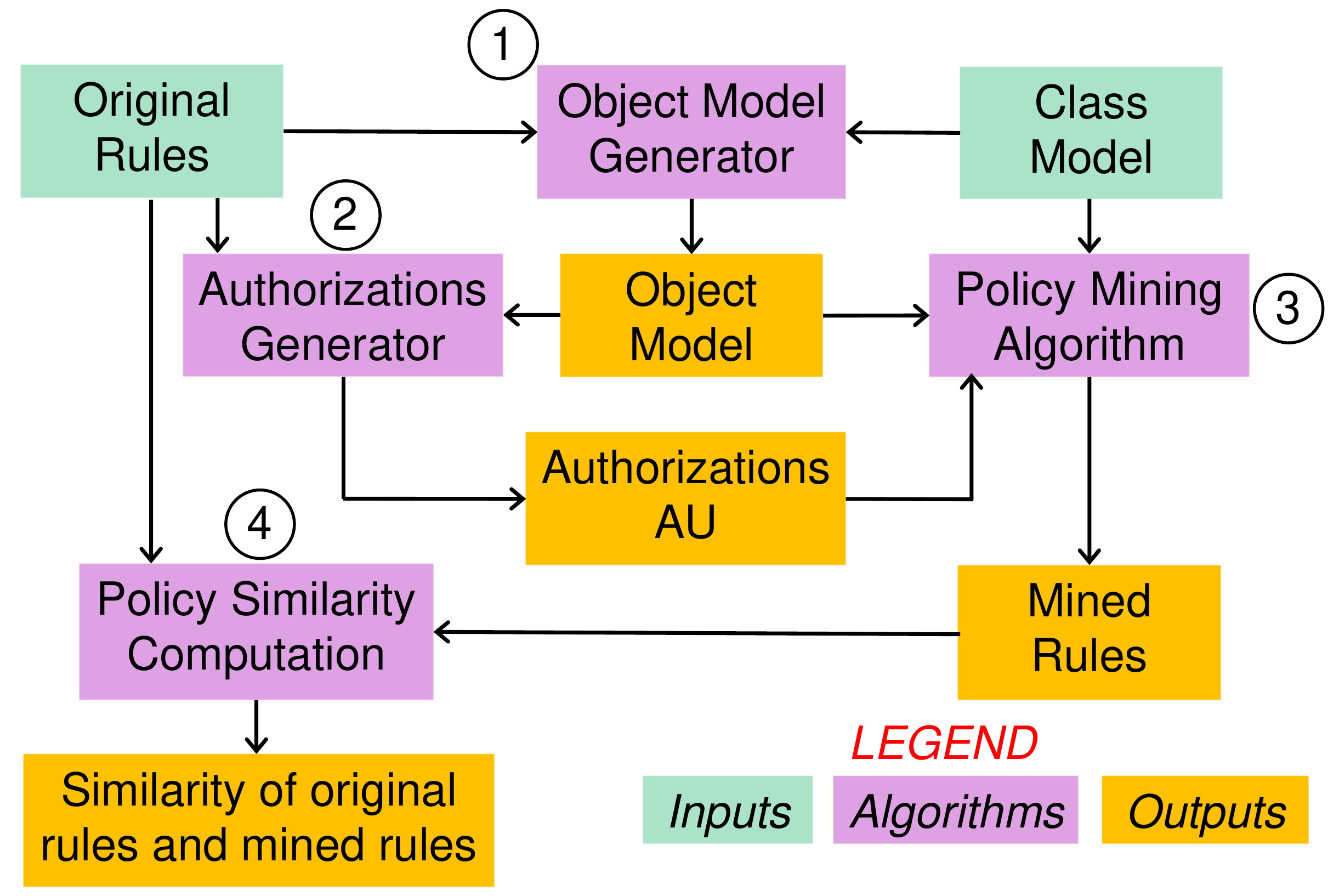}
  \caption{Evaluation methodology.}
  \label{fig:eval-method}
\end{figure}

Our methodology for evaluating policy mining algorithms is depicted in Figure \ref{fig:eval-method}.  It takes a class model and a set of ReBAC rules as inputs. We generate an object model based on the class model (independent of the ReBAC rules), compute the authorizations $\spa$ from the object model and the rules, run the policy mining algorithm with the class model, object model, and $\spa$ as inputs, and finally compare the mined policy rules with the original (input) policy rules.  If the mined rules are similar to the input rules, the policy mining algorithm succeeded in discovering the desired ReBAC rules that are implicit in $\spa$.

To compare FS-SEA* with Bui et al.'s EA \cite{bui19mining}, we use the two ReBAC case study policies in \cite{bui19mining}, which are available at \cite{fssea-software}.  The class models and rules are based on policies of real organizations; the object models are synthetic.  The e-document case study, based on \cite{decat14edocShort}, is for a SaaS multi-tenant electronic-document processing application.  The workforce management case study, based on \cite{decat14workforceShort}, is for a SaaS workforce management application provided by a company that handles the workflow planning and supply management for product or service appointments.  The only change we make is to omit from the workforce management case study the classes and 7 rules related to work orders, because they involve inheritance, which our algorithm does not yet support (it is future work). 

The policies from \cite{bui19mining} do not use the additional constraint operators in our policy language.  Therefore, we developed a pseudorandom synthetic policy generator that produces policies that use these operators (as well as the original ones), and we also use the synthetic policies for evaluation.  Synthetic policies also have the advantage that their size and complexity are easily controlled.  To make the evaluation results more meaningful, our synthetic policy generator is carefully designed to produce policies that have realistic structure, statistically similar in some ways to the realistic sample policies and case studies in \cite{bui19mining}. 

\subsection{Synthetic Policy Generation}
\label{sec:synthetic-policy}



\myparagraph{Class model}


\begin{figure*}[tbp]
\begin{tabular}[t]{@{}l@{}}
  \centering
\includegraphics[width=0.8\textwidth]{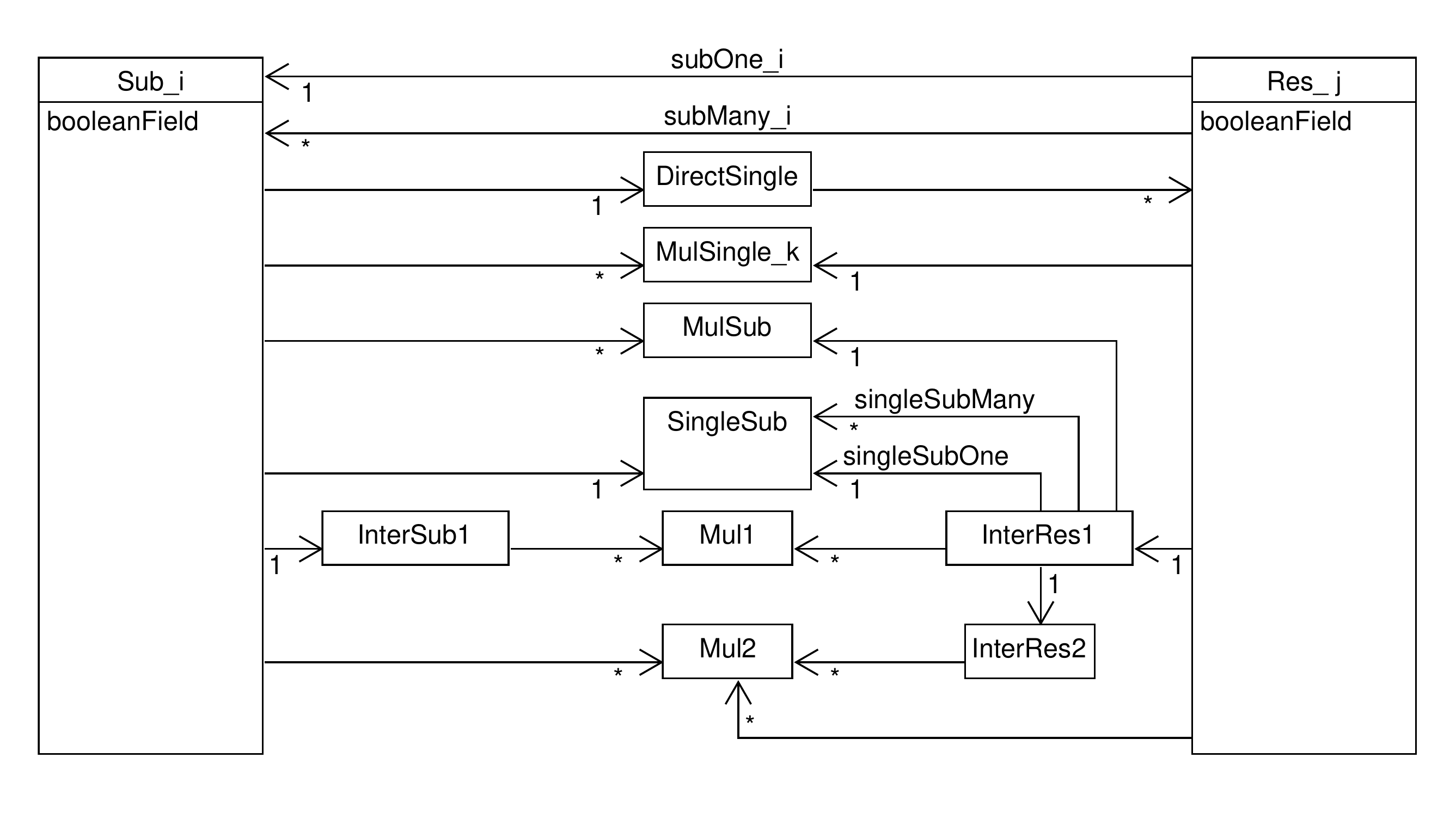}
\end{tabular}
  \caption{Class model for synthetic policies.  Only one subject type Sub\_i and one resource type Res\_j are shown, but the full class model contains five of each, i.e., i=1..5.  Similarly, there are two MulSingle\_k classes, i.e., k=1..2.  Note that each resource class Res\_j has two associations, with different multiplicities, to each subject class Sub\_i.  Names of some associations are omitted to avoid clutter; each of these associations is implicitly named after the target class, e.g., associations pointing to Mul2 are named "mul2".}
  \label{fig:syn-policy-class-model}
\end{figure*}


The generated class model has the structure shown in Figure \ref{fig:syn-policy-class-model}. The class model allows generating atomic conditions and constraints with many combinations of path length(s) and operator; we refer to such a combination as a \textit{condition/constraint type}.  The class model is designed to support the condition/constraint types used in the sample policies and case studies in \cite{bui19mining} (except for a few rarely used condition/constraint types), plus condition/constraint types involving our new constraint operators.  In particular, it supports 10 constraint types and 3 condition types. 
For example, association subOne\_i is used to generate atomic constraints with subject path length 1, resource path length 2, and operator ``equal''; subMany\_i is used to generate similar constraints, except with operator ``in''. 
As another example, DirectSingle class is used to generate atomic constraints with subject path length 3, resource path length 1, and operator ``contains'', and to generate atomic conditions with path length 2 and operator ``in''.

\myparagraph{Object model generation}

The object model generator is parameterized by $N_{sub}$, the desired number of instances of each subject class.  The number of instances of each resource class is $5 \cdot N_{sub}$.  
The numbers of instances of other classes is fixed at 3.  This reflects a typical structure of realistic policies, in which the numbers of instances of some classes (e.g., doctors, patients, health records) scale linearly with the overall size of the organization, while the numbers of instances of other classes (e.g., departments, medical specialties) grow much more slowly (which we approximate as constant).

The values of Boolean fields are chosen randomly.  The values of other fields are randomly chosen object(s) of the appropriate type.  For a field with type $C$ and multiplicity ``many'', the number of chosen objects is randomly chosen to be 1, $|C|-1$, or $|C|$, where $|C|$ is the total number of instances of class $C$.

\myparagraph{Rule generation}

Rule generation uses several numbers and statistical distributions based on the sample policies and case studies in \cite{bui19mining}. The number of rules in each policy is $N_r=20$, which is the average number of rules in those policies.  The rule generator pseudorandomly chooses a subject type and a resource type using uniform distributions, and then picks a number $n_r$ of rules to generate for that pair of subject-resource types.  $n_r$ is chosen to be 1, 2, 3 or 4 with probability of 0.82, 0.12, 0.03, or 0.03, respectively, based on the frequency distribution in the sample policies and case studies. The rule generator then chooses the total number of features (atomic conditions and atomic constraints) for each rule to be 1, 2, or 3 with probability 0.5, 0.25, or 0.25, respectively, based on the frequency distribution in the sample policies and case studies.  For each feature, the generator chooses a condition/constraint type, again based on the frequency distribution in the sample policies and case studies, and then chooses an atomic condition/constraint of that type using a uniform distribution. These steps are repeated until $N_r$ rules have been generated.  Finally, we apply $\simplifyrulesL$ (see Section \ref{sec:evol-alg}) to the generated rules, since hand-written policies typically don't contain unnecessarily complicated rules.




\subsection{Policy Similarity Metrics}
\label{sec:policy-sim-metrics}

We evaluate the quality of the generated policy primarily by its {\em syntactic similarity} and {\em per-rule semantic similarity} to the original policy.  These metrics are defined in \cite{bui19mining} and are normalized to range from 0 (completely different) to 1 (identical). 

\myparagraph{Syntactic Similarity}

Syntactic similarity measures the fraction of types, atomic conditions, atomic constraints, and actions that rules or policies have in common. The Jaccard similarity of sets is $J(S_1, S_2) = |S_1\intersect S_2| \,/\, |S_1 \union S_2|$.  The {\em syntactic similarity of rules} $\rho_1=\langle st_1, sc_1, rt_1,$ $rc_1, c_1, A_1\rangle$ and $\rho_2=\tuple{st_2, sc_2, rt_2, rc_2, c_2, A_2}$ is the average of $J(\set{st_1}, \set{st_2})$, $J(sc_1, sc_2)$, $J(\set{rt_1}, \set{rt_2})$, $J(rc_1, rc_2)$, $J(c_1, c_2)$ and $J(A_1, A_2)$.  The {\em syntactic similarity of rule sets} $\Rules_1$ and $\Rules_2$, {\em SynSim}($\Rules_1$, $\Rules_2$), is the average, over rules $\rho$ in $\Rules_1$, of the syntactic similarity between $\rho$ and the most similar rule in $\Rules_2$. 

\myparagraph{Semantic Similarity}

{\em Semantic similarity} measures the fraction of authorizations that rules or policies have in common.  The {\em semantic similarity of rules} $\rho_1$ and $\rho_2$ is $J(\mean{\rho_1},\mean{\rho_2})$.  We extend this to {\em per-rule semantic similarity of policies} in exactly the same way that syntactic similarity of rules is extended to syntactic similarity of policies.  Note that this metric measures similarity of the meanings of the rules in the policies, not similarity of the overall meanings of the policies (in our experiments, the original and mined policies always have exactly the same overall meaning).

\section{Evaluation Results}
\label{sec:evaluation-results}

SEA is implemented in Java.  Feature selection is implemented in Python using the \href{https://pytorch.org/}{PyTorch} deep learning platform. Experiments were run on Windows 10 on an Intel i7-6770HQ CPU.  Our code and data are available at  \cite{fssea-software}. For the SEA algorithm, we use the same parameter values as in \cite{bui19mining}.  For our feature selection algorithm, we take the limit on the number of training iterations to be  $N_{tr}=10000$, and we set $F_u$ (the fraction of features to be selected as useful) to be 15\% for FS-SEA1 and $5\%$ for each feature selection step in FS-SEA*.
We take the number of neurons in the hidden layer to be 64, based on experiments with a few synthetic policies showing that this is sufficient to learn perfect classifiers for those policies, while using 32 hidden nodes led to a few mis-classifications.



\subsection{Experiments Comparing FS-SEA* with EA}

As described in section \ref{sec:evaluation-methodology}, we use the e-document and workforce management case studies to compare FS-SEA* and EA.  For each case study, we run both algorithms on 5 object models from \cite{bui19mining} and average the results. For e-document, the algorithms achieve the same average syntactic similarity (89\%), and FS-SEA* achieves 6\% higher average per-rule semantic similarity (91\% vs. 85\%). For workforce management, EA achieves 2\% higher syntactic similarity (96\% vs. 94\%), and 1\% higher per-rule semantic similarity (98\% vs. 97\%).  We conclude that FS-SEA* and EA are comparably effective at discovering the desired rules.

The running time of FS-SEA* to run the experiments is higher than EA.  This is because, for the smaller search spaces in these case studies ($|\spa|$ is 2687 for e-document and 1739 for workforce \cite[Table 1]{bui19mining}), the overhead of repeated feature selection and repeated evolutionary search outweighs the potential speedup it can provide.  For the larger synthetic policies ($|\spa|$ up to 23124) used in the experiments in the next section, FS-SEA* is significantly faster than SEA. This does not show the advantage of FS-SEA* in terms of running time since search spaces in evolutionary steps of the case studies are small enough for EA to mine high quality policies. FS-SEA* spends large amount of time on the FS step trying to reduce the search space, and repeat the FS-SEA step in multiple iterations because the algorithm only includes 1 best rule for each tuple of subject type, resource type and action in each iteration. There are several tuples that have more than 1 rule in the input policies of the case studies. 

To show the importance of a sufficiently expressive policy language with the operators needed to express a policy in a natural way, we also run both algorithms on ACLs generated from a synthetic policy with $N_{sub} = 10$ and WSC = 145.  Since EA does not support the operators added to the policy language in this paper, it cannot discover the original policy, and using policy similarity to evaluate the mined policy seems unfair, so we use WSC.  We find that EA produces a much larger policy (WSC = 737) than FS-SEA* (WSC = 144) and takes much longer (155 minutes vs. 5 minutes), because EA must generate many low-quality rules to cover the permissions covered by input rules that use the additional operators.

\subsection{Experiments Comparing FS-SEA* with FS-SEA1 and SEA}

\begin{figure*}[htb]
  \includegraphics[width=0.5\textwidth]{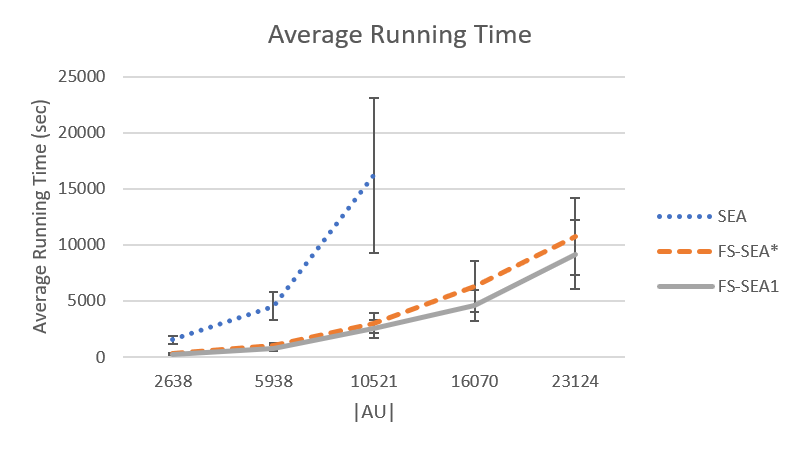}\hspace*{0pt}
  \includegraphics[width=0.5\textwidth]{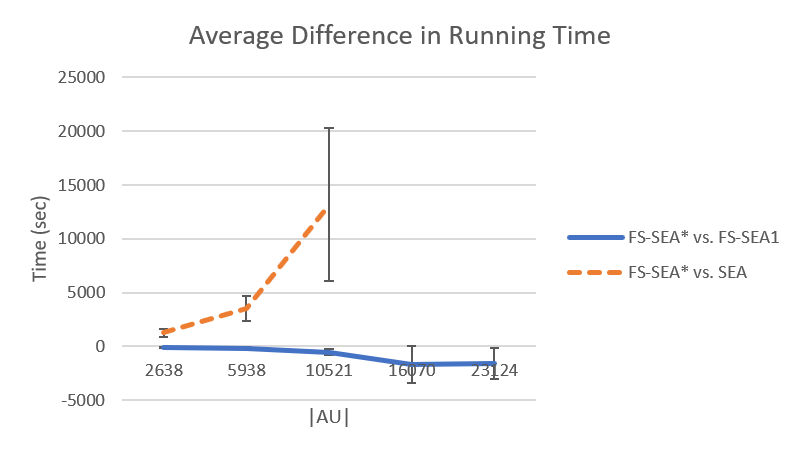}\hspace*{0pt}
  \caption{Left: Running time, in seconds, as a function of |$\spa$|.  Right: Differences between running times, in seconds, as a function of |$\spa$|.}
  \label{fig:running-time}
\end{figure*}

\begin{figure*}[htb]
  \includegraphics[width=0.5\textwidth]{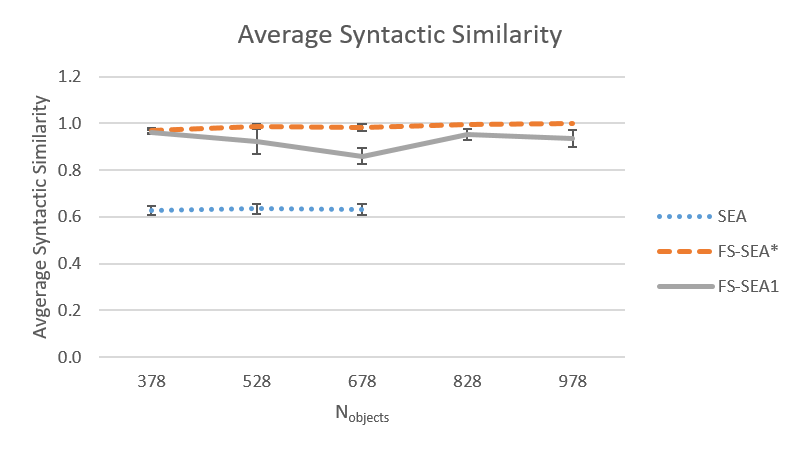}\hspace*{0pt}
  \includegraphics[width=0.5\textwidth]{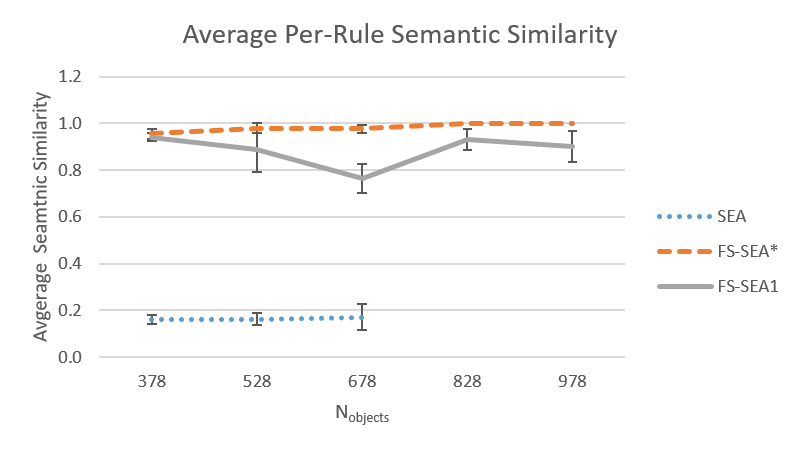}\hspace*{0pt}
  \caption{Syntactic similarity and per-rule semantic similarity over policies of same size as a function of number of objects.}
  \label{fig:general-similarities}
\end{figure*}

\begin{figure*}[htb]
  \includegraphics[width=0.5\textwidth]{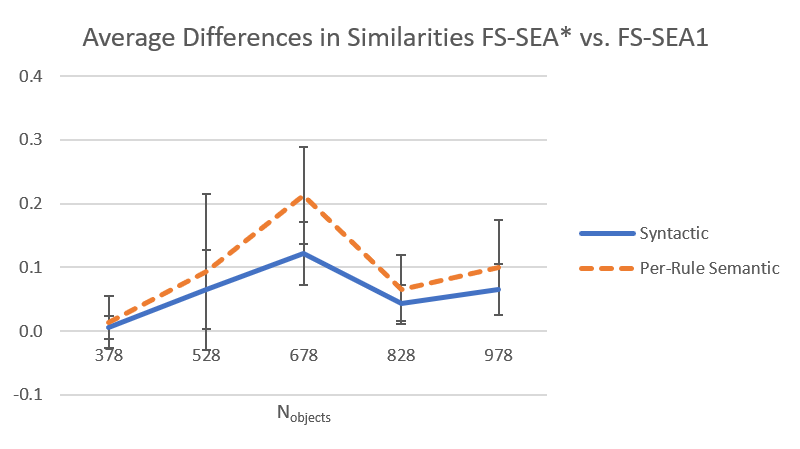}\hspace*{0pt}
  \includegraphics[width=0.5\textwidth]{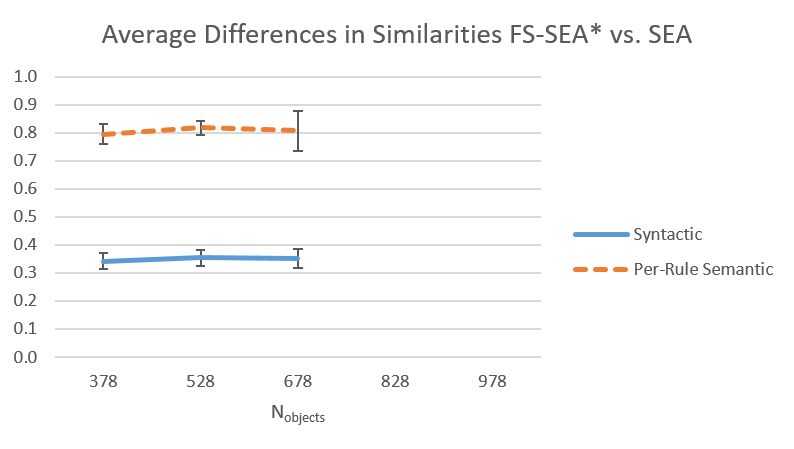}\hspace*{0pt}
  \caption{Differences of syntactic similarity and per-rule semantic similarity between FS-SEA1, SEA and FS-SEA*.}
  \label{fig:diff-similarities}
\end{figure*}

\begin{figure*}[htb]
  \includegraphics[width=0.5\textwidth]{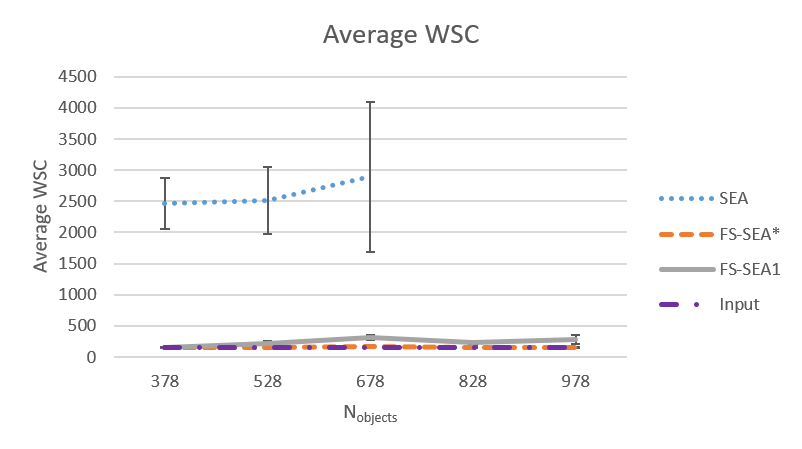}\hspace*{0pt}
  \includegraphics[width=0.5\textwidth]{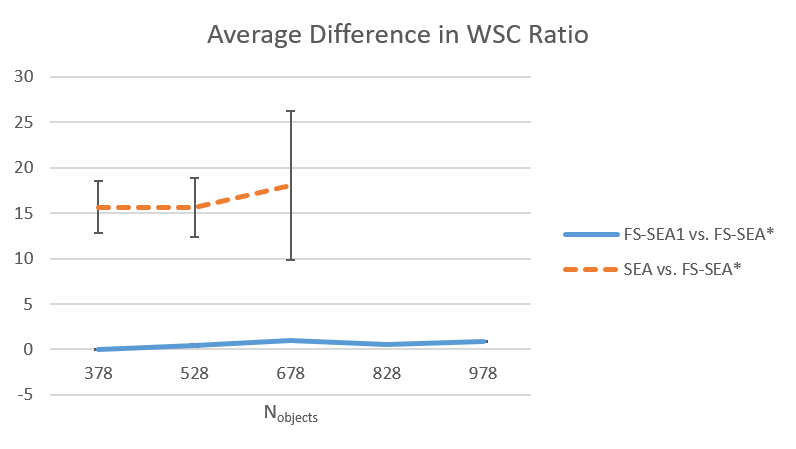}\hspace*{0pt}
  \caption{Left: WSC over policies of same size as a function of number of objects. Right: Differences of WSC Ratio between FS-SEA1, SEA and FS-SEA*.}
  \label{fig:wsc}
\end{figure*}

We generate 5 sets of synthetic rules.  For each of them, we generate a sequence of synthetic object models of varying size, with $N_{sub} = {10, 15, 20, 25, 30}$.  We run all 3 algorithms on all of them.  Each reported data point is the average over the 5 policies with the same $N_{sub}$.  We compare the results for FS-SEA* and SEA to show the benefits of (iterated) FS; \textbf{we find that FS-SEA* (and FS-SEA1) yield significantly better results than SEA for all metrics: running time, policy similarity (including syntactic similarity and per-rule semantic similarity), and WSC. }  


We compare results for FS-SEA* and FS-SEA1 to show the benefits of iteration; we find that FS-SEA* yields better results than FS-SEA1 for policy similarity and WSC, at the expense of slightly higher running time.

We show the comparisons in two kinds of graphs.  One kind contains 3 curves, one for each algorithm. The other kind contains two curves, one for the difference between FS-SEA* and SEA, and one for the difference between FS-SEA* and FS-SEA1.  Error bars (too small to see in some cases) show the standard deviations. The standard deviations are high in some graphs.  While our use of randomized algorithms (namely, evolutionary search) contributes to this, the more important contributing factors are that a small number of samples (namely, 5) was used for each data point in our experiments, and that the 5 synthetic rule sets used for these 5 samples are not especially similar to each other (our synthetic rule generation algorithm allows a high degree of random variation).
 
The x-coordinate in the graphs is the total number of objects in the object model, denoted $N_{objects}$, except in graphs of running time, for which the x-coordinate is the average number of authorizations $|\spa|$, because we expect running time to be more strongly correlated with $|\spa|$.   



\myparagraph{Running time}

The graphs in Figure \ref{fig:running-time} show that feature selection reduces running time: both algorithms that include FS are significantly faster than SEA, especially for larger policies.  The graphs also show that FS-SEA* takes slightly longer than FS-SEA1, due to the time needed to repeat feature selection and evolutionary search.  The graphs lack results for SEA for the two largest policy sizes, because SEA was so slow that we aborted the experiments.  We aborted execution of SEA on the first object model with $N_{sub} = 25$ after 23 hours; for comparison, FS-SEA* took about 2.7 hours on that object model.

The main reason that SEA is slower than FS-SEA* is that, without feature selection to narrow down the search space, the evolutionary search produces rules with lower quality, each typically covering fewer SRA-tuples in $\spa$.  Therefore, SEA needs to generate more rules than FS-SEA* in order to cover all of $\spa$.  This is costly, because each rule requires two evolutionary searches: one to generate it, and another to try to improve it.


\myparagraph{Policy similarity}

The graphs in Figure \ref{fig:general-similarities} show that feature selection improves policy similarity.  SEA yields average syntactic similarity of about 0.6 and average per-rule semantic similarity of about 0.2, while FS-SEA* and FS-SEA1 yield average syntactic similarity and per-rule semantic similarity in the range of 0.8 to 1.0.  Figure \ref{fig:diff-similarities} (right) shows that FS-SEA* yields an improvement of about 0.3 in syntactic similarity and an improvement of about 0.8 in per-rule semantic similarity, compared to SEA.  Figure \ref{fig:diff-similarities} (left) shows that FS-SEA* achieves slightly better to modestly better results than FS-SEA1 for both policy similarity metrics for all policy sizes.

\myparagraph{WSC}

The graph in Figure \ref{fig:wsc} (left) shows that feature selection improves WSC.  SEA produces policies with WSC roughly in the range 2500 to 3000, while FS-SEA* produces policies with WSC very close to the WSC of the input, roughly in the range 140 to 210, and FS-SEA1 produces policies with WSC somewhat higher but still below 400.  Figure \ref{fig:wsc} (right) shows the average differences in WSC ratio, which is the ratio of the WSC of the mined policy to the WSC of the input policy, for FS-SEA1 and SEA compared to FS-SEA*.  We graph the difference in the WSC ratios, rather than the difference in the WSCs, since the former should be less dependent on policy size.  This graph emphasizes that FS-SEA* produces the best results for all policy sizes (the improvement is negligible but  positive even for the smallest policy size in the graph).



\section{Related Work}
\label{sec:related}

\myparagraph{Related work on ReBAC policy mining}

The only prior work on mining of ReBAC policies (or object-oriented ABAC policies with path expressions) is by Bui et al.  Section \ref{sec:intro} discusses the relationship of our work to their ReBAC mining algorithms in \cite{bui17mining,bui19mining}.   They also developed variants of their algorithms for mining ReBAC policies from incomplete and noisy information about granted permissions \cite{bui18mining}, which may typically be obtained from access logs.  Their modifications to EA to handle these issues can easily be incorporated in FS-SEA*.

\myparagraph{Related work on ABAC Policy mining}

Xu and Stoller proposed the first algorithm for ABAC policy mining \cite{xu15miningABACShort}, a greedy algorithm that is the basis for Bui et al.'s greedy algorithm for ReBAC policy mining, discussed in Section \ref{sec:intro}.  Medvet et al. pioneered the use of evolutionary algorithms for ABAC policy mining \cite{medvet2015}; their work inspired Bui et al.'s evolutionary algorithm for ReBAC policy mining \cite{bui19mining}, which our work extends and improves.    Iyer et al. developed the first ABAC policy mining algorithm that can mine ABAC policies containing DENY rules as well as PERMIT rules \cite{iyer2018}.  Their algorithm is relatively expensive: the worst-case complexity is $O(N_E^2 N_A^5)$, where $N_E$ is the total number of well-formed SRA-tuples (whether permitted or denied), and $N_A$ is the number of attributes. Extending our algorithm to support mining DENY rules is future work.

A few papers have considered the problem of mining ABAC policies from incomplete (and in some cases noisy) information about granted permissions; such information is typically obtained from access logs, so this variant of the problem is often called ``mining from logs''.  Xu and Stoller developed a variant of their greedy algorithm for mining from logs \cite{xu14miningABAClogs}.  Mocanu et al. \cite{mocanu2015} proposed a different approach that learns a Restricted Boltzmann Machine (RBM) by training on the logs and constructs candidate rules by sampling from the RBM, and then constructs a policy from the candidate rules.  Their paper presents only preliminary results from the first phase of their algorithm on one small ABAC policy; the last phase of the algorithm was not implemented.

Cotrini et al. propose a different formulation of the problem of mining from logs and an algorithm, called Rhapsody, to solve it \cite{sparselogs2018}.  Rhapsody is based on APRIORI-SD, a machine-learning algorithm for subgroup discovery.  Rhapsody's running time is sensitive to the number of predicates (conditions and constraints, in our terminology) that can appear in rules.   Rhapsody works well when this number is relatively small, but is already much slower than Xu et al.'s algorithm \cite{xu14miningABAClogs} for an ABAC policy involving only 16 attributes (the number of predicates grows with the number of attributes) \cite{bui18mining}.  Rhapsody can easily be extended to handle path expressions and therefore to support a form of ReBAC policy mining.  However, the number of predicates is much larger with ReBAC than ABAC ({\em cf.} discussion of state space size in Section \ref{sec:intro}), and Rhapsody's running time would be impractical except on small problem instances.

A top-down approach to ABAC policy mining has also been pursued, with the goal of using natural language processing and machine learning to extract ABAC policies from natural language documents. Since this problem is extremely difficult, the focus so far has been on sub-problems, such as analyzing natural language documents to identify the sentences relevant to access control \cite{narouei2017} and the relevant attributes \cite{alohaly2018}.


\svonly{
\begin{acknowledgements}
\thanksText
\end{acknowledgements}}


%
\ieeeonly{\bibliographystyle{IEEEtran}}\acmonly{\bibliographystyle{ACM-Reference-Format}}\lncsonly{\bibliographystyle{splncs04}}\articleonly{\bibliographystyle{alpha}}\svonly{\bibliographystyle{plain}}
\bibliography{references}


 \end{document}